\documentclass[epj]{svjour}
\usepackage{graphics}
\usepackage[ansinew]{inputenc}
\begin{document}
\title{Stability of two-component alkali clusters formed on helium nanodroplets}

\author{G. Droppelmann\inst{1} \and M. Mudrich\inst{2} \and C.P. Schulz\inst{3} \and F.
Stienkemeier\inst{2}} \institute{Fakult\"at f\"ur Physik, Universit\"at Bielefeld, 33615
Bielefeld, Germany \and Physikalisches Institut, Universit\"at Freiburg, 79104 Freiburg,
Germany \and Max-Born-Institut, Berlin, Germany}
\date{Received: date / Revised version: date}

\abstract
{The stability of two-component clusters consisting of light (Na or K) and heavy
(Rb or Cs) alkali atoms formed on helium nanodroplets is studied by femtosecond laser
ionization in combination with mass spectrometry. Characteristic stability patterns
reflecting electron shell-closures are observed in dependence of the \textit{total}
number of atoms contained in the mixed clusters. Faster decay of the stability of mixed
clusters compared to the pure light ones as a function of size indicates a destabilizing
effect of heavy alkali atoms on light alkali clusters, presumably due to second order
spin-orbit interaction.
\PACS{
      {PACS-key}{discribing text of that key}   \and
      {PACS-key}{discribing text of that key}
     } 
} 
\maketitle
%


\section{Introduction}

Metal clusters have proven to be
particularly well suited test objects for studying the transition from molecular quantum
dynamics to solid state physics~\cite{Heer:1993,Haberland:1994,Ekardt:1999,Tiggesbaumker:2007}.
Clusters formed of monovalent alkali metals have been studied in great detail both
experimentally as well as theoretically~\cite{Heer:1993,Haberland:1994,Ekardt:1999,Visser:2003,Brack:1993,Issendorff:2005}.
In these simple metal clusters the electronic structure is dominated by the number of
valence electrons whereas the ionic cores are of secondary importance. The electrons are
delocalized, and the electronic system exhibits a shell structure that is closely related
to the well-known nuclear shell structure. The simple model of the free-electron gas inside a spherical potential well of the dimension of the cluster (Jellium model) applies particularly well to alkali clusters~\cite{Haberland:1994,Ekardt:1999,Brack:1993,Issendorff:2005}.

Important information on the electronic structure of metal clusters has been gained from
simple cluster abundance spectra, reflecting the stability with respect to fragmentation:
Clusters in which the number of valence electrons matches the spherical shell-closing
numbers are produced more abundantly. In addition, odd-even alternations reflect the
enhanced stability of paired electron configurations~\cite{Knight:1984,Haberland:1994}.
These shell effects have also been observed with metal clusters formed in helium nanodroplets~\cite{Diederich:2005}.

Besides the strongly bound clusters (covalent or metallic) which are usually observed in experiments,
alkalis can form van der Waals-type molecules and clusters in which all electrons are
spin-oriented and strongly localized~\cite{Visser:2003,Schulz:2004}. Although bonding in these high-spin clusters was
predicted to be quite strong for lithium clusters (`ferromagnetic
bonding'), sodium clusters were theoretically found to be much more loosely bound than their metallic
counterparts~\cite{Visser:2003}, and the same behavior may be expected for the heavier
alkali species.

While most atomic species reside inside the helium droplets due to the attractive
interaction with the helium surroundings, alkali atoms and molecules are weakly bound to
the droplet surface in bubble states. This weak binding energy of
the order of 10\,K leads to the fact that out of all clusters formed on the droplet surface
preferentially the weakly bound ones remain attached to the droplets~\cite{Stienkemeier:1995}.
This leads to an enrichment of weakly bound high-spin
diatomic and triatomic molecules of up to a factor 10$^4$~\cite{Higgins:1998,Nagl:2008}.

The formation of high-spin alkali clusters using the helium nanodroplet technique has been reported by our group~\cite{Schulz:2004}. In this previous experiment we observed characteristic differences in the abundance spectra
of light and heavy alkali clusters. While alkali clusters of sizes up
to 25 atoms were seen in the case of sodium (Na) and potassium (K), cluster sizes
exceeding 5 and 3 atoms are strongly suppressed in the case of rubidium (Rb) and cesium
(Cs), respectively. This behavior has been interpreted in terms of the reduced stability
of high-spin states of Rb and Cs clusters with respect to depolarization, leaving behind
hot, unpolarized clusters that may fragment and escape out of the detection volume.
Depolarization may be induced by the strong second-order spin-orbit interaction present
in the heavy alkali atoms, causing spontaneous spin flipping into the unpolarized state.
Alternatively, the localized, spin-oriented electrons may evolve into a delocalized
collective state as the cluster size grows larger, which eventually gives rise to spin flipping~\cite{Schulz:2004}.


Based on these findings the question arises how stable mixed clusters of light and heavy
alkali atoms in high-spin states are. In other words, how does the binding of one or several heavy alkali
atoms (Rb or Cs) to a cluster consisting of light alkalis (Na or K) affect the stability
of the compound cluster? In order to shed some light on this issue we report on a series
of measurements of abundance spectra of mixed alkali clusters formed in high-spin states
on the surface of helium nanodroplets using femtosecond (fs) photo ionization (PI) in combination
with mass-selective ion detection.

\section{Experimental}

Besides numerous techniques for producing beams of metal clusters~\cite{Heer:1993}, the
aggregation of metal atoms inside helium nanodroplets has been established as an
alternative route to forming metal clusters of well defined
composition~\cite{Tiggesbaumker:2007}. Using this technique, the metal clusters are
formed in the ultracold environment of the helium droplets at temperatures in the
millikelvin range.
Moreover, helium nanodroplets can be efficiently
loaded with a variety of different atomic or molecular species.


In the experiment reported here, a beam of helium nanodroplets is consecutively doped
with two different species of alkali atoms in two separate pickup cells to form two-component clusters on the
droplets. The growth statistics for alkali clusters is found to deviate from
the Poissonian distribution~\cite{Vongehr:2003}. Further downstream the mixed clusters are photo ionized by fs laser
pulses from a mode-locked Ti:Sa laser. Nonresonant fs PI at high pulse repetition rate (80\,MHz) is an efficient method of ionizing alkali clusters largely size-independently. The experimental details are given
elsewhere~\cite{Droppelmann:2005,Claas:2006,Claas:2007}.


PI takes place inside the detection volume of a commercial quadrupole mass
spectrometer, in which the laser beam intersects the doped helium droplet beam
perpendicularly. The output of the fs laser is tightly focused into the helium
droplet beam.
The laser wavelength is tuned to about 760\,nm, which corresponds to the
shortest possible wave length. These laser parameters are chosen on account of the fact
that PI signals recorded at the alkali cluster masses generally increase
with shorter wave length and with tighter focusing of the laser beam, thereby mostly
suppressing mass-specific resonances of the ionization cross sections.


\section{Results and discussion}

\begin{figure}
\center\resizebox{1.0\columnwidth}{!}{\includegraphics{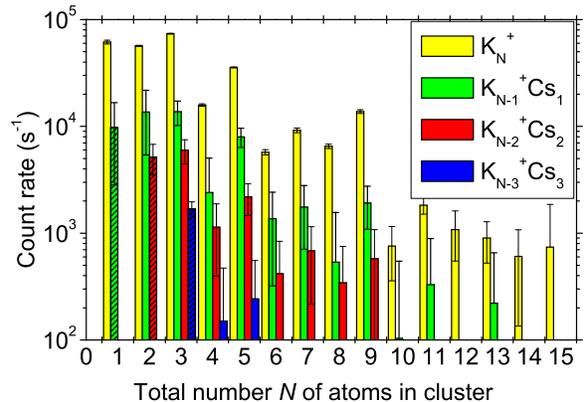}}
\caption{Photo ionization mass spectra of two-component potassium-cesium clusters
K$_{N-M}$Cs$_{M}$ for $M=0$ to $M=3$ formed on helium nanodroplets. The pure cesium clusters
are represented by hatched bars. } \label{fig:CsK}
\end{figure}
\begin{figure}
\center\resizebox{1.0\columnwidth}{!}{\includegraphics{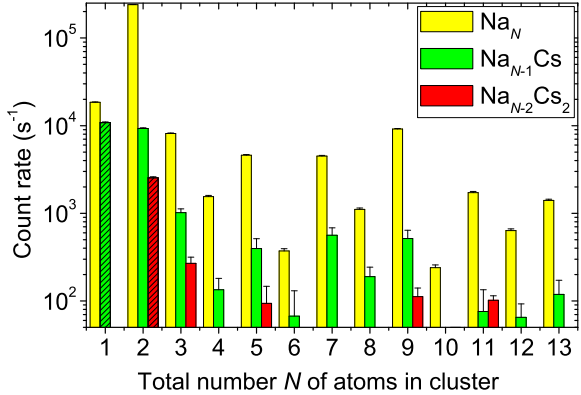}}
\caption{Mass spectra of two-component sodium-cesium clusters Na$_{N-M}$Cs$_{M}$ for
$M=0$ to $M=2$ formed on helium nanodroplets.} \label{fig:NaCs}
\end{figure}

When multiply doping helium nanodroplets simultaneously with Cs and either
Na or K atoms, both single-component as well as two-component clusters appear in the mass
spectrum, as shown in Fig.~\ref{fig:CsK} and Fig.~\ref{fig:NaCs}. As abscissa we choose
the total number $N$ of atoms contained in each cluster to reveal similarities in the
abundance patterns of pure and mixed clusters.

The mass spectrum of mixed K$_{N-M}$Cs$_M$ clusters for $M=0-3$ shown in
Fig.~\ref{fig:CsK} is recorded when doping helium droplets with an average of 11 Cs
atoms in a first pickup cell at a temperature of 105\,$^{\circ}$C and then with 13 K
atoms in a second pickup cell at a temperature of 160\,$^{\circ}$C. The error bars
reflect uncertainties resulting from background count rates caused by the following
effects~\cite{Droppelmann:2005}: Even when helium droplets are doped with only one species, non-negligible signal
intensity is measured at cluster masses matching the ones of mixed clusters. This is due
to the fact that natural potassium has two stable isotopes $^{39}$K and $^{41}$K with
abundances 93.3\,\% and 6.7\,\%, respectively. Thus, K$_N$ clusters are composed of all
possible combinations of K isotopes $^{39}$K$_M$\,$^{41}$K$_N$, the abundances of which
being given by the binomial distribution. Besides, contamination of the droplets mainly
with water on the level of 1\,\% leads to the appearance of alkali clusters with water molecules attached to
them, K$_N$H$_2$O, or even to reaction products of the type K$_N$(KOH). However, no influence on cluster growth has ever been observed.


The mixed clusters, K$_{N-1}$Cs$_1$, and K$_{N-2}$Cs$_2$ plotted in Fig.~\ref{fig:CsK}
as green and red bars, respectively, display a strikingly similar abundance pattern
compared to the one of pure K$_N$ clusters. The excess energy due to the nonresonant nature of
the ionization process leads to desorption and fragmentation of the ionized clusters.
Thus, the abundance spectra reflect the stability pattern with respect to fragmentation
rather than the abundance upon formation on the helium droplets. Prominent steps at
masses 3, 5 and 9 clearly reflect the electronic shell closures of the \textit{ionized}
clusters. Besides, the mass spectra display a pronounced odd-even
alternation with odd-numbered clusters being more abundant. Thus, the stability of
two-component K$_{N-M}$Cs$_M$ clusters with respect to the electronic structure appears
to be determined by the \textit{total} number of constituents and not by the number of
atoms of one species. Since under the given experimental
conditions fragmentation of strongly bound metallic clusters would not be expected, we
conclude that mixed as well as pure alkali clusters formed on helium nanodroplets are in high-spin
states~\cite{Schulz:2004}.

Interestingly, mixed clusters containing one or two Cs atoms tend to fall off
slightly faster with increasing number of K atoms than pure K$_N$ clusters. This trend is
even more pronounced as the number of Cs atoms grows larger. This points at the fact that
mixed K-Cs clusters tend to be destabilized by the presence of Cs atoms, in particular
when more than one Cs atom is involved. Mixed clusters containing one Cs atom,
K$_{N-1}$Cs$_1$, are detected up to $N=13$, those containing two Cs atoms,
K$_{N-2}$Cs$_2$, are observed up to $N=9$, and K$_{N-3}$Cs$_3$ clusters are detected with
high uncertainty up to $N=5$.

The peak intensities of pure Cs$_N$ clusters are significantly lower than the ones of
pure K$_N$ clusters. Since no substantial difference in the doping efficiencies is
expected, this discrepancy is attributed to a more efficient PI process in
the case of K$_N$ clusters. In contrast, two-component clusters, K$_{N-M}$Cs$_M$, are
expected to have PI cross sections comparable to those of pure K$_N$
clusters. The fact that mixed clusters K$_{N-1}$Cs$_1$ and K$_{N-2}$Cs$_2$ are
significantly less abundant than pure K$_N$ clusters therefore implies that these mixed
clusters actually are less stable.

\begin{figure}
\center\resizebox{1.0\columnwidth}{!}{\includegraphics{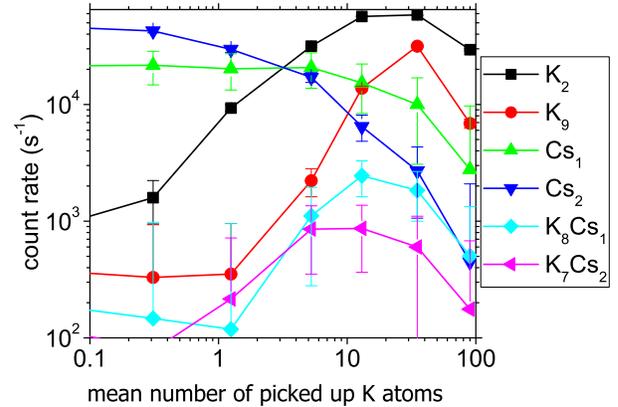}}
\caption{Evolution of integral signal intensities of selected cluster mass peaks as a
function of the degree of doping with potassium atoms. Cesium doping conditions are held
constant at an average of 11 cesium atoms per droplet. } \label{fig:CsKDoping}
\end{figure}

Besides the mentioned stability pattern as a function of the number of K atoms, the more
pronounced limitation to maximum cluster size appears to be imposed by the number of Cs
atoms attached to a mixed clusters of a given size. E.\,g., mixed clusters containing three
Cs atoms, K$_{N-3}$Cs$_3$, are less abundant by about one order of magnitude compared to
mixed clusters with one Cs, K$_{N-1}$Cs$_1$, which roughly matches the reduced abundance
of the pure Cs$_3$ cluster as compared to the Cs monomer (hatched bars in
Fig.~\ref{fig:CsK}). However, this behavior is a consequence of co-doping the helium
droplets with K atoms. When loading the droplets only with Cs atoms at the same pickup
cell temperature, the detected Cs$_N$ abundance spectrum is peaked at Cs$_2$ with roughly
equal signal intensities at the masses of Cs and Cs$_3$.

In order to study this effect a series of measurements at variable K pickup conditions is
depicted in Fig.~\ref{fig:CsKDoping}. Shown are count rates of a representative selection
of different mass peaks as a function of the approximate mean number of picked up K
atoms. These values are computed from the vapor pressure curve of K using the calibration
10$^{-4}$\,mbar $\widehat{\approx}$ 1 atom per droplet. This calibration has been checked
by means of laser-induced fluorescence spectroscopy in earlier experiments.

The data shown in Fig.~\ref{fig:CsKDoping} indicate the following general
trend. Both pure Cs$_1$ and Cs$_2$ clusters as well as combined clusters K$_{8}$Cs$_1$
and K$_{7}$Cs$_2$ are suppressed in the mass spectrum upon multiple doping with K atoms.
In particular, abundances of clusters containing two Cs atoms, Cs$_2$ and K$_{7}$Cs$_2$,
fall off more rapidly than the ones containing only one Cs atom. When adding up all
signals from clusters containing one, respectively two Cs atoms, (not shown in the graph)
the same trend holds. Thus, the rapid disappearance of mixed clusters of a given size
containing more than one Cs atom is a combined effect of the low stability of large pure
Cs$_N$ clusters and the reduction of stability of mixed clusters with an increasing
number of K atoms.

Fig.~\ref{fig:NaCs} displays mass spectra of two-component clusters containing Na and Cs
atoms. The pure Na$_N$ cluster abundances are dominated by the Na$_2$ dimer mass peak due to a resonance in the
PI cross section of Na$_2$ at the chosen laser wave
length~\cite{Claas:2007}. Mixed clusters composed of Na and Cs atoms are less
abundant than pure Na$_N$ clusters by about one order of magnitude. This is mainly due to
a lower level of doping with Cs atoms compared to the experiment with mixed clusters made
of K and Cs, discussed above. Thus, combinations involving 2 Cs atoms, Na$_{N-2}$Cs$_2$,
are only visible at selected cluster sizes and Na$_{N-3}$Cs$_3$ is not present in the
spectrum. Nevertheless, a faster droping off of the mixed cluster intensities of
Na$_{N-1}$Cs$_1$ with respect to the pure Na$_N$ mass peaks is again clearly visible, indicating that Na$_N$
clusters tend to be destabilized by the presence of Cs atoms as it is seen in the spectra
of mixed clusters with K atoms.

In addition to the mixed alkali clusters discussed so far, clusters made up of Rb and either Na or K have been studied. In comparison with the K$_{N-M}$Cs$_M$ cluster spectrum, larger quantities of mixed
clusters K$_{N-M}$Rb$_M$ are observed. Clusters containing up to $M=5$ Rb atoms are clearly
detected. The abundance pattern both of pure an mixed clusters as a function of total
number of atoms, $N$, is modulated by the effect of shell closures and odd-even
alternations, as in the K-Cs case. Also, mixed cluster abundances fall off faster as $N$
grows larger compared to pure K$_N$ clusters, indicating the destabilizing effect of
additional heavy alkali atoms. The abundance of mixed clusters of a fixed size is also
reduced the more Rb atoms are present in the clusters. However, the latter effects are not
as pronounced as in the K-Cs combination. Two-component clusters containing only one Rb
atoms, K$_{N-1}$Rb$_1$, appear systematically less abundant than those containing
more Rb atoms. Furthermore, the influence of the order of doping of first K atoms in a
first pickup cell and then Rb atoms in a second one, and vice versa, was also compared.
However, no significant difference in the cluster abundance spectra was observed.

The mass spectrum of mixed Na$_{N-M}$Rb$_M$ clusters is similar to the one of K$_{N-M}$Rb$_M$ except for the fact that pure Na$_N$ clusters have much lower abundances compared to pure Rb$_N$ clusters and even
compared to mixed clusters, which is due to the small ionization cross section of Na$_N$
for $N>2$. As in the K$_{N-M}$Rb$_M$ mass spectra, clusters
containing only one Rb atom, Na$_{N-1}$Rb$_1$, are systematically less abundant than
those containing a larger number of Rb atoms. This type of mixed clusters apparently has
minimum stability as compared to all other cluster compositions. Cluster sizes as large
as Na$_2$Rb$_8$ are observed in the mass spectrum.

\section{Conclusion}

Two-component high-spin alkali clusters composed of light (Na, K) and heavy (Rb, Cs) atoms can be
formed on helium nanodroplets with rather large abundance. The stability pattern of the
mixed clusters with respect to photo fragmentation follows the one of pure clusters, with
the \textit{total} number of atoms in the clusters being the determining parameter.
Characteristic features such as electron-shell closures and odd-even alternations are
clearly visible in the mass spectra. No sudden breakdown of cluster abundance upon
addition of one or a specific number of heavy alkali atoms to the cluster is observed.
However, the abundance of mixed clusters of a given size is systematically lower than the
one of the pure lighter alkali clusters Na$_N$ and K$_N$ of the same size, roughly
following the dropping abundance distributions of pure heavy alkali clusters Rb$_N$ and
Cs$_N$. Furthermore, for a fixed number of Rb or Cs atoms attached to Na$_N$ of K$_N$
clusters, the abundance of the mixed clusters falls off faster as the number of Na or K
atoms increases than for the pure Na$_N$ of K$_N$ clusters. Thus, admixing heavy Rb or Cs
atoms to light Na$_N$ of K$_N$ clusters tends to destabilize the resulting two-component
clusters.

The fact that the number of heavy alkali constituents imposes the dominant limitation to
the maximum size of stable two-component clusters rather than the total number of atoms
points at second-order spin-orbit interaction being the dominant effect that leads to
depolarization and fragmentation. Delocalization of the electron wave functions which
would cause spontaneous depolarization as the high-spin clusters grow larger can
therefore be ruled out. The largest two-component clusters are observed in the
combination of Na and Rb yielding mixed clusters of sizes up to Na$_{10}$Rb$_6$ and
Na$_2$Rb$_8$. No significant influence of the doping order on the mass spectra of mixed
clusters, as studied with K-Rb clusters, is observed.

However, the helium droplet source conditions are found to crucially influence the
abundance of pure and mixed clusters. Larger droplets have higher cooling capacity and
therefore sustain the aggregation of larger spin-polarized clusters and possibly of
metallic clusters as well. Therefore studying abundance spectra of pure and mixed alkali
clusters at variable helium droplet beam conditions may provide valuable information on
the formation and stability of high-spin clusters. Besides, time-resolved measurements of
the fragmentation process, e.g. using the pump-probe technique, will add to the
understanding of the laser induced dynamics of many-body systems isolated in helium
nanodroplets.

%
%
%

Financial support by the Deutsche Forschungsgemeinschaft is gratefully acknowledged.

\bibliographystyle{unsrt}

\end{document}